\documentclass{ws-ijmpe}
\newcommand{\LamS}{$\Lambda(1405)$ }
\newcommand{\SigS}{$\Sigma^{0}(1385)$ }

\begin{document}

\markboth{I.~Zychor}{Excited Hyperons Produced in pp Collisions with ANKE@COSY}

\catchline{}{}{}{}{}

\title{EXCITED HYPERONS PRODUCED IN PROTON-PROTON COLLISIONS WITH ANKE AT COSY\\
}

\author{\footnotesize IZABELLA ZYCHOR
}

\address{Instytut Problem\'ow J\c{a}drowych, Pl-05-400 \'Swierk, Poland\\
i.zychor@fz-juelich.de}

\maketitle

\begin{history}
\received{(received date)}
\revised{(revised date)}
\end{history}

\begin{abstract}
Excited neutral hyperons $Y^{0*}$ produced in the \mbox{$pp \rightarrow pK^+ Y^{0*}$} reaction with a COSY 
beam momentum of 3.65 MeV/c have masses below 1540~MeV/c$^2$. 
The ANKE spectrometer allows the simultaneous observation
of different decay modes: $Y^{0*} \rightarrow \pi^0 \Sigma^0, \pi^\mp \Sigma^\pm, 
\pi^0 \Lambda, K^- p$ 
by measuring kaons and pions of either charge in coincidence with protons.
\ \\
We have found indications for a neutral excited hyperon resonance $Y^{0*}$ 
with a mass of $M(Y^{0*}) = (1480\pm 15)~\textrm{MeV/c}^2$ and
a width of $\Gamma(Y^{0*}) = (60\pm 15)~\textrm{MeV/c}^2$.
The cross section for $Y^{0*}$ is of the order of few
hundred nanobarns.
It can be either a $\Sigma^0$ or a $\Lambda$ hyperon
and on the basis of existing data no conclusion could be made whether it is a
three--quark baryon or an exotic state.
\ \\
Missing- and invariant--mass techniques have been used to identify the $\Lambda(1405)$ resonance
decaying via $\Sigma^0\pi^0$. 
The cross section for $\Lambda(1405)$ production is equal to 
$(4.5 \pm 0.9_{\rm{stat}}\pm 1.8_{\rm{syst}})\,\mu\textrm{b}$.
The shape and position of the $\Lambda(1405)$ distribution
are similar to those found from other decay modes, so no support is given to
the two-pole model.
\end{abstract}

\section{Introduction}

The production and properties of hyperons have been studied for more
than 50 years, mostly in pion and kaon induced reactions.
Hyperon production in \textit{pp} collisions has been investigated close to
threshold at SATURNE (Saclay, France) and COSY-J\"ulich.
Reasonably
complete information on $\Lambda(1116)$, $\Sigma^0(1192)$,
$\Sigma^0(1385)$, $\Lambda(1405)$ and $\Lambda(1520)$ can be found in
the Review of Particle Physics.\cite{PDG}

For the $\Lambda(1405)$, in spite
of rather high statistics achieved (the total world statistics is
several thousand events), there are still open questions concerning
the nature of this resonance: is it a singlet \textit{qqq} state in
the frame of SU(3) or a quark-gluon \textit{(uds-q)} hybrid, or a KN
bound state?\cite{Loring}$^-$\cite{Jido2003}

On the contrary, the $\Sigma(1480)$ hyperon is not well established yet
and it is described as a 'bump' with unknown quantum numbers.\cite{PDG}
The Crystal Ball experiment has not seen any indications for the resonance $\Sigma(1480)$
in the $\pi^0 \Lambda$ invariant mass distribution measured in the reaction 
$K^-p \rightarrow \pi^0 \pi^0 \Lambda$, dominated
by the $\Sigma(1385)$.\cite{Crystal_PRC69_042202}

The program to investigate hyperon production from \textit{pp} interactions at low energies is 
very well suited for the ANKE spectrometer operated at COSY--J\"ulich.

\begin{figure}[h]
\centerline{\psfig{file=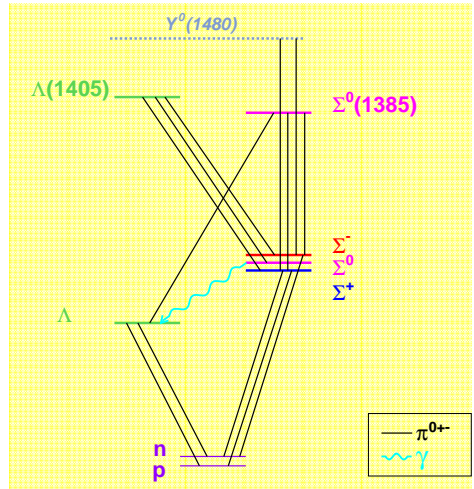,height=6.5cm}}
\caption{Simplified decay scheme of investigated hyperons.}
\label{fig:rozpady}
\end{figure}

In Fig.~\ref{fig:rozpady} a simplified decay scheme of hyperons investigated in our
experiments is presented.

\section{Experiment and particle identification with ANKE}

The experiments have been performed with the ANKE spectrometer\cite{ANKE_NIM}
at the Cooler SYnchrotron COSY at the Research Center J\"ulich (Germany).\cite{COSY}

COSY is a medium energy cooler synchrotron and storage ring for 
both polarized and unpolarized protons and deuterons.
At COSY various targets can be used, \textit{e.g.} solid or cluster--jet.
COSY provides beams in the momentum range between 0.6 and 3.7 GeV/c.
 
\begin{figure}[ht]
\centerline{\psfig{file=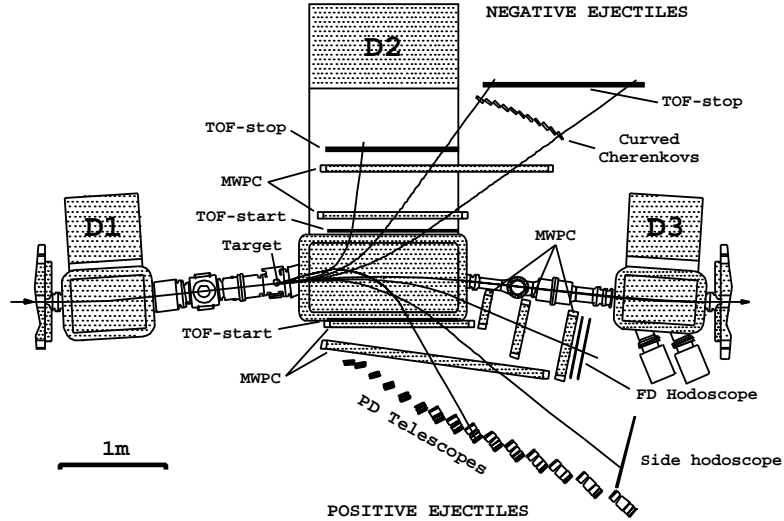,height=7.0cm}}
\caption{ANKE spectrometer and detectors.}
\label{fig:ANKE}
\end{figure}

ANKE ("\textbf{A}pparatus for Studies of \textbf{N}ucleon and \textbf{K}aon 
\textbf{E}jectiles") is a magnetic spectrometer 
located at an internal target position of COSY.
It consists of three dipole magnets, see Fig.~\ref{fig:ANKE}. 
The central C--shaped spectrometer dipole D2,
placed downstream of the target, separates the reaction products from
the circulating COSY beam.  The ANKE detection system, comprising
range telescopes, scintillation counters and multi--wire proportional
chambers, simultaneously registers both positively and negatively
charged particles and measures their momenta.\cite{K_NIM}

The ANKE telescopes are used to register positively charged particles.
They discriminate pions, kaons and protons with 
the same momenta due to their different energy losses. Passive
copper degraders in the telescopes between the scintillation
counters enhance the discrimination efficiency.
The $K^+$~mesons are stopped in the $\Delta E$ counters or in 
the second degrader of each telescope.
Their decay, mainly into
$\mu^+\nu_\mu$ and $\pi^+\pi^0$ with a lifetime of $\tau = 12.4$~ns,
provides a very effective criterion for kaon identification via
detection of delayed signals in a so-called veto counter (with
respect to prompt signals from \textit{e.g.}\ a $\pi^+$ produced in
the target passing over all counters of a telescope).
By measuring such delayed signals from the decay of stopped kaons, positively charged 
kaons can be identified at ANKE in a background of pions, protons and 
scattered particles up to 10$^6$ times more intense.\cite{EPJ_A22_301}
The use of veto counters causes a decrease of particle identification efficiency, 
typically by factor of 6.
The tracks of the ejectiles, measured with multi-wire
proportional chambers (MWPCs), are used to reconstruct momenta of any registered particle.

Data originally taken for scalar meson\cite{MB_scalar} and $\phi$\cite{MH_phi} production 
have been used to study 
the production of low-lying hyperon resonances in \textit{pp} collisions
with ANKE@COSY. Experiments have been performed in 2002 and 2005, respectively.

In Table~\ref{tab:exp_details} details of the experimental conditions are given.
\begin{table}[b]
\tbl{Experimental details.}
{\begin{tabular}{@{}lcc@{}} \toprule
          \hphantom{00}            & \hphantom{0}4 weeks in 2002 & \hphantom{0}4 weeks in 2005 \\ 
                                   & \hphantom{0}for Y$^{0*}(1480)$& \hphantom{0} for $\Lambda(1405)$ \\ \colrule 
integrated luminosity\hphantom{00} & \hphantom{0}6 pb$^{-1}$                    & \hphantom{0}70 pb$^{-1}$  \\
coincidences\hphantom{0}           & 3 particles (p, K$^+$, $\pi^+$ or $\pi^-$)\hspace*{1cm} & 4 particles (p, p, K$^+$, $\pi^+$) \\
$K^+$ / $\pi^+$ momentum                  & \multicolumn{2}{c}{\hspace*{0.6cm}0.2 - 0.9 GeV/c}\\
p momentum                  & \multicolumn{2}{c}{  \hspace*{0.6cm}$>$ 0.75 GeV/c}\\
$\pi^-$ momentum                  & 0.4 - 1.0 GeV/c &0.2 - 1.0 GeV/c\\
delayed veto for $K^+$\hphantom{0} & yes                                        & no\hphantom{0} \\
detection efficiency\hphantom{0}   & 7\%                                        & 55\%\hphantom{0} \\
mass resolution                    & $\sim$ 10 MeV/c$^2$                             & $\sim$ 20 MeV/c$^2$  \hphantom{0} \\ \botrule
\end{tabular}}
\label{tab:exp_details}
\end{table}

\section{Excited neutral hyperon \textit{Y$^{0*}$}(1480)}

Final states comprising a proton, a positively charged kaon, a pion
of either charge and an unidentified residue X were investigated in the
reaction $pp \rightarrow p{K}^+Y \rightarrow p{K}^+\pi^\pm X^\mp$ at the beam momentum of 3.65~GeV/c.  
Kaons are identified by measuring delayed signals from the their decay, 
which, together with well-defined pions and protons, are used to determine the mass of~X. 

\begin{figure}[h]
\hspace*{-0.5cm}{\psfig{file=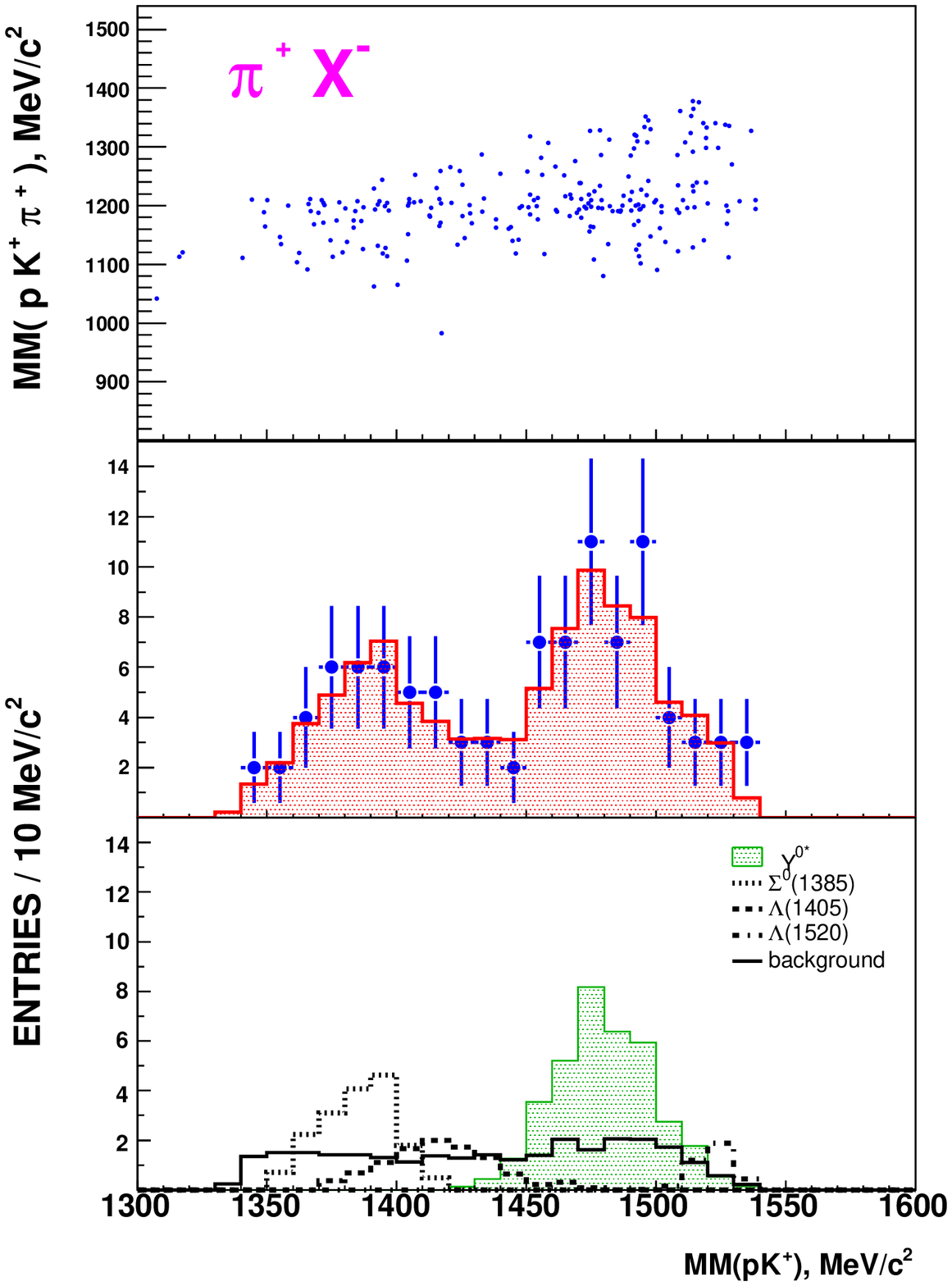,height=8.0cm}}\hspace*{0cm}{\psfig{file=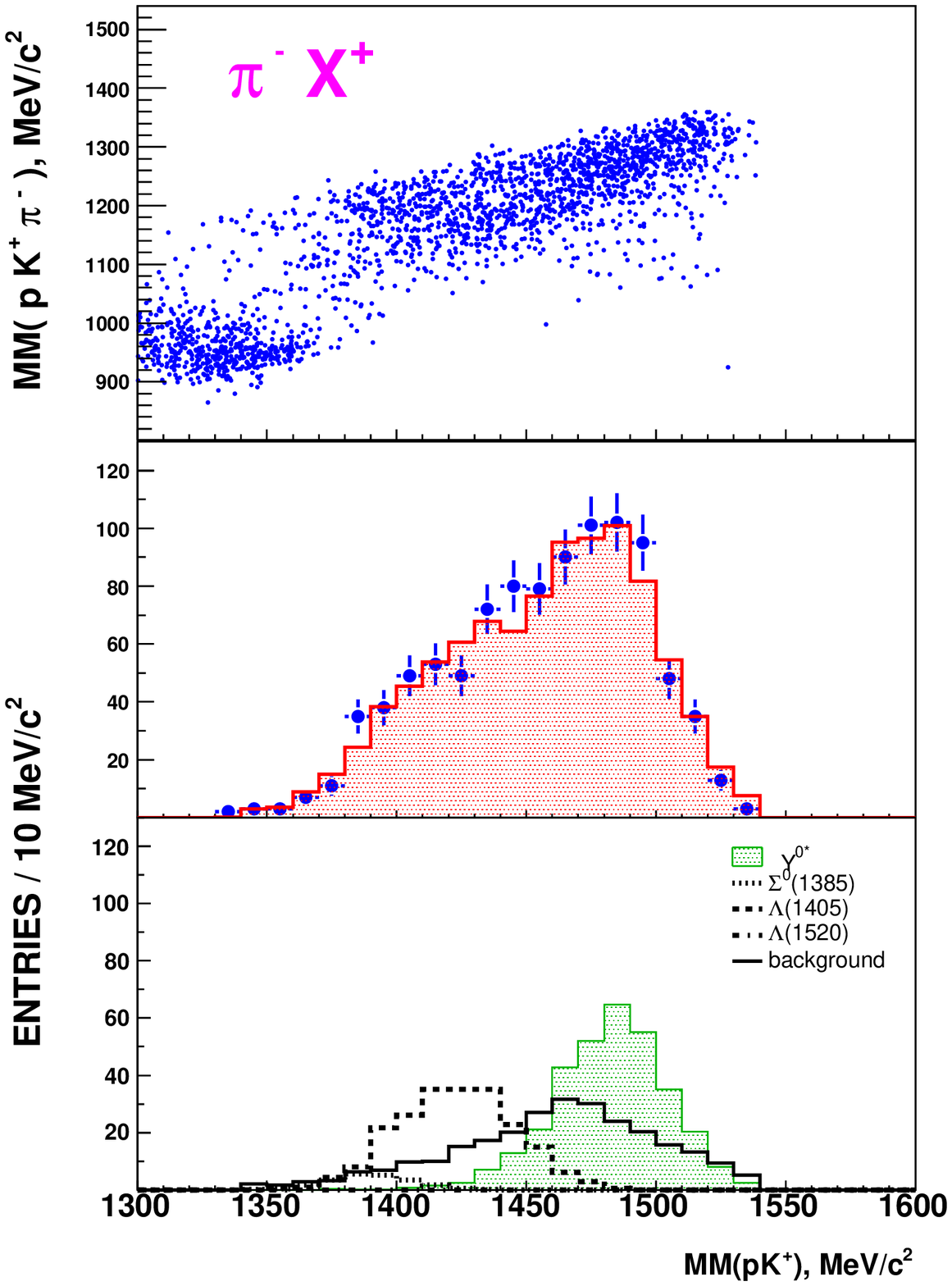,height=8.0cm}}
\caption{Missing--mass $MM(p K^+)$ spectra for the reaction \mbox{$pp \rightarrow
pK^+\pi^+X^-$}(left) and \mbox{$pp \rightarrow pK^+\pi^-X^+$}(right).
Upper parts: $MM(pK^+)$ \textit{versus} $MM(pK^+\pi)$ for $\pi^+$ (right) and $\pi^-$ (left).
Middle parts: Comparison of experimental (points) and simulated (shaded histograms) distributions.
Lower parts: Non--resonant and resonant contributions to the overall simulated   
histograms.} 
\label{fig:pos_neg_1480}
\end{figure}

In the upper part of Fig.~\ref{fig:pos_neg_1480} the missing mass distributions $MM(pK^+\pi)$
\textit{versus} $MM(pK^+)$ 
are shown for the reactions \mbox{$pp \rightarrow
pK^+\pi^+X^-$}(left) and \mbox{$pp \rightarrow pK^+\pi^-X^+$}(right). 
Since the probability for
detecting three--particle coincidences $(pK^+\pi^+)$ is about an
order of magnitude smaller than for $(pK^+\pi^-)$, the resulting
numbers of events are also drastically different.
In the distribution for the reaction $pp\to p{K}^+\pi^+X^-$(left) 
an enhancement corresponding to $X^-=\Sigma^-(1197)$, is observed on
a low background. 
In the charge--mirrored \mbox{$pp \rightarrow pK^+\pi^-X^+$} case(right), the
$\pi^-$ may originate from different sources, \textit{e.g.}\ a decay
with the $\Sigma^+(1189)$ or a secondary decay of $\Lambda \to
p\pi^-$, arising from the major background reaction $pp \rightarrow
pK^+\Lambda \rightarrow pK^+\pi^-p$. Protons from this reaction are
easily rejected by cutting the missing mass $MM(pK^+\pi^-)$ around the proton mass.

If only events around the $\Sigma$ mass are selected, then the missing
mass spectrum $MM(pK^+)$ in the reaction $pp\to p {K}^+ \pi^+ X^-$
shows two peaks, see in the middle--left part in Fig.~\ref{fig:pos_neg_1480}. One of them
corresponds to the contribution of $\Sigma^0(1385)$ and
$\Lambda(1405)$ hyperons. The second peak is located at a mass
$\sim$1480~MeV/c$^2$. In the $\pi^-X^+$ case, the distribution also
peaks at 1480~MeV/c$^2$, see the right--middle part in Fig.~\ref{fig:pos_neg_1480}.

We have assumed that the measured missing mass $MM(pK^+)$ spectra can
be explained by the production of hyperon resonances and non-resonant
contributions.  Detailed Monte Carlo simulations have been performed
including the production of well established excited hyperons~
($\Sigma^0$(1385), $\Lambda$(1405), $\Lambda$(1520)) and
non-resonant contributions like $pp \rightarrow NK^+\pi X$ and $pp
\rightarrow NK^+\pi\pi X$; $X$ denotes any hyperon which could be
produced in the experiment.  
For both final states the shape of the measured distributions cannot 
be reproduced by the simulations and an excess of events is observed around
the missing mass of $1480~\textrm{MeV/c}^2$.
Thus, an the excited hyperon $Y^{0*}$ decaying via $\pi^\pm X^\mp$ 
was included into simulations. 
The best fit to the experimental data was obtained for the $Y^{0*}$ with a mass
$M(Y^{0*})= (1480\pm15)~\textrm{MeV/c}^2$ and a width $\it\Gamma$$(Y^{0*}) =
(60\pm 15)~\textrm{MeV/c}^2$, see lower parts in Fig.~\ref{fig:pos_neg_1480}.
There have been identified 100 and 1000 events for $\pi^+ X^-$ and  $\pi^- X^+$ case,
respectively.
The statistical significance of the signal,
assuming that this is due to the production of the $Y^{0*}$,
is between 4 and 6~$\sigma$ depending on a procedure.
The production cross section is of the order of few hundred nanobarns.\cite{1480}

\section{The $\Lambda(1405)$ hyperon}

The \mbox{$pp \rightarrow pK^+ p \pi^- X^0$} reaction is selected by 
a multiparticle final state, containing two protons, a positively
charged kaon, a negatively charged pion and an unidentified
residue $X^0$.
In the $\Sigma^0(1385) \rightarrow \Lambda \pi^0$
decay the $X^0$ residue is a $\pi^0$ while, for the $\Lambda(1405)
\rightarrow \Sigma^0 \pi^0$ decay, $X^0 = \pi^0 \gamma$ (see
Fig.~\ref{fig:rozpady}).
\begin{figure}[ht]
\centerline{\psfig{file=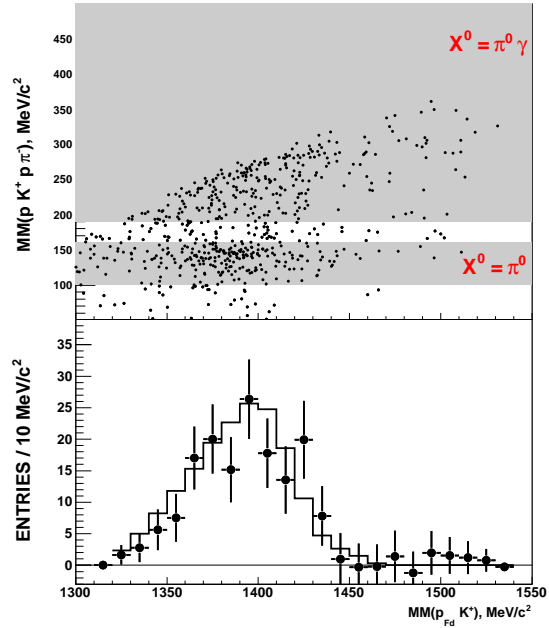,height=10.0cm}}
\caption{
Upper panel: Missing mass $MM(pK^+\pi^- p)$ \textit{versus} $MM(p_{Fd}K^+)$ with
the shaded horizontal boxes showing the $MM(p K^+\pi^- p)$ bands used
for event selection. The lower one is located around the~$\pi^0$~mass
and the upper one selects $MM(p K^+\pi^- p) >$ 190\,MeV/$c^2$,
significantly greater than the~$\pi^0$~mass.
Lower panel: The background--subtracted line shape of
the $\Lambda(1405)$ decaying into $\Sigma^0 \pi^0$ (points) compared
to the spectrum of the $\Lambda(1405)$ from the LEPS experiment for
photon energy range of $1.5 < E_{\gamma} < 2.0$ GeV (solid line).
} 
\label{fig:Lambda_1405}
\end{figure}
In the upper part of Fig.~\ref{fig:Lambda_1405}  
a distribution of $MM(p_{Fd}K^+)$ \textit{versus} $MM(p K^+\pi^-
p)$ is plotted for events with the invariant mass $M(p_{Sd}\pi^-)$ 
of the $p_{Sd} \pi^-$ pairs corresponding to the mass of the $\Lambda$, 
\textit{i.e.}~between 1112 and 1120\,MeV/c$^2$. 
The two horizontal bands show the four--particle missing--mass
$MM(p K^+\pi^- p)$ criteria used to separate the $\Sigma^0(1385)$
candidates from those of the $\Lambda(1405)$. The lower band is
optimised to identify a $\pi^0$ whereas the upper one selects
masses significantly greater than $m(\pi^0)$. 

In order to extract the $\Lambda(1405)$ distribution from the
measured $\Sigma^0 \pi^0$ decay, 
the non--resonant contributions have been fitted to the experimental data. 
After subtracting them from the data, the distribution shown as experimental
points in the lower panel of Fig.~\ref{fig:Lambda_1405} was obtained. 156 events have been 
identified in this spectrum.\cite{1405}

The $(\Sigma \pi)^0$ invariant--mass distributions have been
previously studied in two hydrogen bubble chamber experiments.
Thomas \textit{et al.} found $\sim$\,400
$\Sigma^+\pi^-$ or $\Sigma^-\pi^+$ events corresponding to the
$\pi^- p \rightarrow K^0 \Lambda(1405)\rightarrow K^0 (\Sigma
\pi)^0$ reaction at a beam momentum of 1.69\,GeV/c.\cite{Thomas}
Hemingway used a 4.2\,GeV/c kaon beam to investigate
$K^- p \rightarrow \Sigma^+(1660) \pi^- \rightarrow \Lambda(1405)
\pi^+ \pi^- \rightarrow  (\Sigma ^+ \pi^-) \pi^+ \pi^-$ and
measured 1106 events.\cite{Hem}
Recently, the LEPS experiment 
has investigated the \LamS hyperon production in the 
$\gamma p \rightarrow K^+ Y^*$ reaction.\cite{LEPS_1385_1405}
The \LamS hyperon was measured in the $(K^+\Sigma^\pm\pi^\mp)$ final state,
where the contamination from \SigS was estimated from the ($K^+\Lambda\pi^0$) final
state.
In the lower panel in Fig.\ref{fig:Lambda_1405} our experimental points are compared 
to the results of the LEPS experiment (for comparison with data of Thomas and Hemingway
see Ref.~19). 
Despite the very different production mechanisms, all four distributions 
have similar shapes and positions.
This might suggest that, if there are two states present in this
region, then the reaction mechanisms in the four cases are
preferentially populating the same one. 
It should, however, be
noted that by identifying a particular reaction mechanism, the
proponents of the two--state solution can describe the shape of
the distribution that we have found.\cite{Geng}

\section{Outlook}

The decay of excited hyperons $Y^{0*}$ \textit{via} $\Lambda
\pi^0$ and $\Sigma^0 \pi^0 \rightarrow \Lambda \gamma \pi^0$ can
be detected directly in electromagnetic calorimeters by
registering neutral particles, \textit{i.e.}\ $\gamma$ and/or
$\pi^0$. Measurements of such channels are discussed for $\gamma
p$~reactions with CB/TAPS~at~ELSA\cite{ELSA} and are also planned in $pp$
collisions with WASA~at~COSY\cite{WASA}

\section*{Acknowledgements}

We thank all
other members of the ANKE collaboration and the COSY accelerator
staff for their help during the data taking. This work has been
supported by COSY-FFE Grant, BMBF, DFG and Russian Academy of
Sciences.

\end{document}